\documentclass[prl,twocolumn,showpacs]{revtex4}

\usepackage{epsfig}
\usepackage{graphics}
\usepackage{amsmath}

\newcommand{\ket}[1]{\left\vert#1\right\rangle}

\newcommand{\Eq}[1]{Eq.(\ref{#1})}

\begin{document}
\author{Yoav Sagi}
\affiliation{Department of Physics, Technion - Israel Inst. of
Tech., Haifa 32000, Israel}
\title{New scheme for generating GHZ like state of $n$ photons }
\begin{abstract}
In this paper we propose a new scheme for creating a three photons
GHZ state using only linear optics elements and single photon
detectors. We furthermore generalize the scheme for producing any
GHZ-like state of $n$ photons. The input state of the scheme
consists of a non-entangled state of $n$ photons. Experimental
aspects regarding the implementation of the scheme are presented.
Finally, the role of such schemes in quantum information
processing with photons is discussed.
\end{abstract}

\maketitle

In recent years considerable attention has been drawn to the role
played by entanglement in quantum physics. In particular,
entanglement was found to contradict local realism
\cite{einstein_35,bell_64}, which lies at the heart of classical
mechanics. Moreover, as the number of particles participating in
the entangled state gets larger, this violation of local realism
becomes even stronger \cite{GHZ_90,kaszlikowski_00}. Entanglement
was also found to be the key ingredient in many quantum
information processing applications, such as dense coding
\cite{bennett_92}, teleportation \cite{bennett_93} and
cryptography \cite{ekert_91}. In this field, one of the main
questions is what is the most suitable physical implementation of
quantum information processing systems. Among several
possibilities, photons and linear optics offer the advantages of
slow decoherence processes and ease of creation and manipulation
in experiments. Nevertheless, they also pose the problem of very
inefficient multi-photons interactions. For example, it was proved
by Calsamiglia and Lutkenhaus \cite{calsamiglia_00} that using
only linear optics, it is impossible to implement a perfect Bell
state analyzer, which also means that it is impossible to create
an entangled state from a non entangled state using only linear
optical elements. This non-interacting nature is problematic in
the implementation of quantum computation, because a deterministic
c-not gate composed of solely linear optical elements cannot
exist. One way to overcome this problem is by letting go of the
requirement for a deterministic gate \cite{knill_01}. A different
approach was suggested in Ref. \cite{raussendorf_00}. In their
paper, the authors propose a different model for quantum
computation, in which the input state is a special entangled state
called ''cluster state'', and all subsequent computational
operations are carried out using only measurements. The usefulness
of their new computational model to an optical implementation
using photons as qubits depends on the ability to create an
initial entangled state. Nevertheless, the benefit of this new
computational model is that once the initial state is created we
do not need to use $2$-qubits operations, and we may exploit this
advantage together with the idea of post-selection, as will be
explained later.

In this letter we present a new scheme to create a GHZ-like
(Greenberger-Horne-Zeilinger) state of $n$ photons
\begin{eqnarray}
\ket{\psi}&=&\frac{1}{\sqrt{2}}\Big{(}\ket{0^o}_{B1}\ket{0^o}_{B2}...\ket{0^o}_{Bn}+\nonumber\\
&&\ket{90^o}_{B1}\ket{90^o}_{B2}...\ket{90^o}_{Bn}\Big{)} \ \ ,
\end{eqnarray}
using only linear optical elements and single photon detectors.
For $n=3$ this state enables us to violate local realism in a
single measurement \cite{GHZ_90}. The first experimental
observation of a GHZ state of 3 photons was reported in Ref.
\cite{bouwmeester_99}. In this experiment, two entangled photon
pairs were used to create a GHZ state. Due to the very small
probability of simultaneous double pair creation in the non-linear
crystal they used in the experiment, the count rates were of the
order of $6\cdot10^{-3}$ counts per second. One of the novel ideas
in the scheme presented here is that the input state is not
entangled. This enables us to suggest a way to implement its
creation in a much more efficient way, and thus the GHZ state rate
of creation is significantly larger. Another advantage of the new
scheme, as will be explained later in detail, is its ability to be
generalized to $n$-photons GHZ-like state. We would also like to
note the work of Zou \textit{et al} \cite{zou_02} who proposed a
scheme for the creation of a W-state. In their suggestion they
also use linear optical elements, single photon detectors and an
input state which consists of two single photons and one entangled
pair. Also should be noted the work of Fiur\'{a}\v{s}ek
\cite{fiurasek_02}, who proposed a scheme for a probabilistic
creation of the entangled state of two optical modes
$\ket{\psi}=1/\sqrt{2}(\ket{N;0}+\ket{0;N})$, where $\ket{M;K}$
denotes $M$ photons in the first mode and $K$ photons in the
second mode.

\begin{figure}
    \begin{center}
    \includegraphics[scale=0.15]{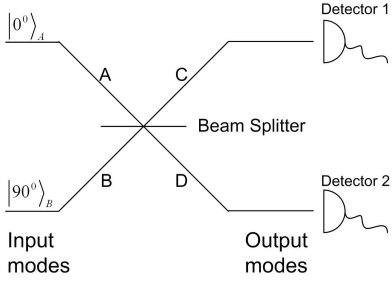}
    \end{center}
    \caption{Scheme for creating $\ket{\psi^-}$ using post-selection.}\label{fig4}
\end{figure}

Before going into the details of the new scheme, we would like to
go back to a well known method of creation of the Bell state
\begin{equation}
\ket{\psi^-}=\frac{1}{\sqrt{2}}\Big{(}\ket{0^o}_A\ket{90^o}_B-\ket{90^o}_A\ket{0^o}_B\Big{)}
\ \ ,
\end{equation}
where $\ket{0^o}_A$ denotes one photon with polarization oriented
at $0^o$ degrees at mode $A$ \cite{ou_88,bose_02}. This setup of
two photons will be an important segment in the new scheme. The
setup is given in figure \ref{fig4}.
The input state of the scheme is the non-entangled state
$\ket{0^o}_A\ket{90^o}_B$. After the beam-splitter the output
state is
\begin{eqnarray}\label{two_photons}
\ket{\psi}_{output}&=&\frac{1}{2}(\ket{0^o}_C+\ket{0^o}_D)\otimes(\ket{90^o}_C-\ket{90^o}_D)\nonumber\\
&=&\frac{1}{2}(\ket{0^o}_D\ket{90^o}_C-\ket{90^o}_D\ket{0^o}_C)+\nonumber\\
&&\frac{1}{2}(\ket{90^o}_C\ket{0^o}_C-\ket{90^o}_D\ket{0^o}_D) \ \
.
\end{eqnarray}
When doing the experiment, half the times the two detectors fire
at once, and half the times only one of them fires. If we restrict
ourselves to the ensemble of experiments in which the two
detectors fired together, we post-select from the state given in
\Eq{two_photons} the Bell state $\ket{\psi^-}$. If one now
introduces polarizers in front of the two detectors, one can carry
out a Bell-inequality experiment which will prove the quantum
nature of the state \cite{ou_88}.

\begin{figure}
    \begin{center}
    \includegraphics[scale=0.23]{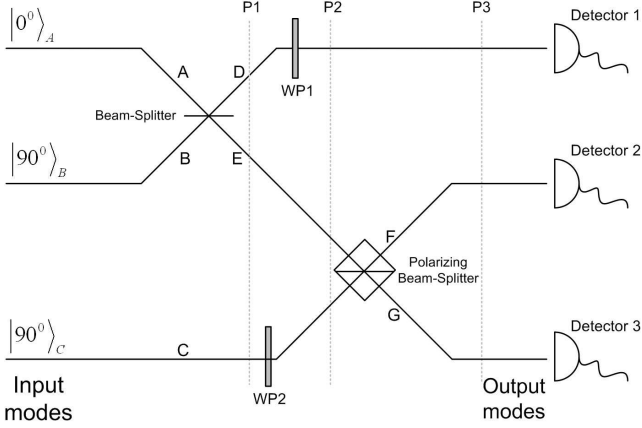}
    \end{center}
    \caption{Scheme for creating a GHZ state.}\label{fig1}
\end{figure}

We shall now turn to describe the scheme for creating a 3-photon
GHZ state. A graphic presentation of it is given in figure
\ref{fig1}.
The input state is
\begin{equation}
\ket{\psi}_{input}=\ket{0^o}_A\ket{90^o}_B\ket{90^o}_C \ \ .
\end{equation}
The first optical element is a beam-splitter which mixes modes $A$
and $B$ in the same way as was described before. The state at
plane $P1$ is given by
\begin{eqnarray}\label{plain_P1}
\ket{\psi}_{P1}&=&\frac{1}{2}\Big{(}
(\ket{0^o}_{E}\ket{90^o}_{D}-\ket{90^o}_{E}\ket{0^o}_{D})\otimes\ket{90^o}_{C}+\nonumber\\
&&(\ket{0^o}_{D}\ket{90^o}_{D}-\ket{0^o}_{E}\ket{90^o}_{E})\otimes\ket{90^o}_{C}
\Big{)} \ \ .
\end{eqnarray}
Between plane $P1$ and $P2$ there are two $\lambda/2$ waveplates.
$WP1$ is placed such that it retardation axis is rotated at $45^o$
relative to polarization axis $90^o$. Thus, the effect of this
waveplate is to transform $\ket{90^o}_D\rightarrow \ket{0^o}_D$
and $\ket{0^o}_D\rightarrow -\ket{90^o}_D$. $WP2$ is placed such
that its retardation axis is rotated at $22.5^o$ relative to
polarization axis $90^o$, thus transforming the photon at mode $C$
$\ket{90^o}_C\rightarrow 1/\sqrt{2}(\ket{0^o}_C+\ket{90^o}_C)$.
The state at plane $P2$ is given by
\begin{eqnarray}\label{plain_P2}
\ket{\psi}_{P2}&=&\frac{1}{2\sqrt{2}}\Big{(}
(\ket{90^o}_{D}\ket{90^o}_{E}+\ket{0^o}_{D}\ket{0^o}_{E})\otimes\nonumber\\
&&(\ket{0^o}_{C}+\ket{90^o}_{C})-\nonumber\\
&&(\ket{0^o}_{E}\ket{90^o}_{E}+\ket{0^o}_{D}\ket{90^o}_{D})\otimes\nonumber\\
&&(\ket{0^o}_{C}+\ket{90^o}_{C}) \Big{)} \ \ .
\end{eqnarray}
The last optical element is a polarizing beam splitter, which we
shall assume lets $\ket{0^o}$ photons pass and reflects
$\ket{90^o}$ photons. Thus, the state at plane $P3$ is given by
\begin{eqnarray}\label{plain_P3}
\ket{\psi}_{P3}&=&\frac{1}{2\sqrt{2}}\Big{(}
(\ket{90^o}_{D}\ket{90^o}_{F}+\ket{0^o}_{D}\ket{0^o}_{G})\otimes\nonumber\\
&&(\ket{0^o}_{F}+\ket{90^o}_{G})-\nonumber\\
&&(\ket{0^o}_{G}\ket{90^o}_{F}+\ket{0^o}_{D}\ket{90^o}_{D})\otimes\nonumber\\
&&(\ket{0^o}_{F}+\ket{90^o}_{G}) \Big{)} \ \ .
\end{eqnarray}
Now let us restrict ourselves to the cases where there is a single
photon in each of the modes $D$,$F$ and $G$. It is easy to verify
that the ensemble of experiments in which this condition is
fulfilled corresponds to projection on the state
\begin{equation}
\ket{\psi}_{PS}=\frac{1}{2\sqrt{2}}(\ket{90^o}_{D}\ket{90^o}_{F}\ket{90^o}_{G}+\ket{0^o}_{D}\ket{0^o}_{F}\ket{0^o}_{G})
\ \ ,
\end{equation}
where the subscript $PS$ stands for \emph{Post-Selected}. This
state is the well known GHZ state. It is not normalized due to the
fact the we obtain the wanted state only $1/4$th of the times we
carry out the experiment. Experimentally, the post-selected
subspace corresponds to the case where all three detectors fire at
once. The projection to this subspace can be achieved using
electronic coincidence logic for the three detectors.

There are some important points to stress regarding the scheme
just presented. First, the photons must arrive at the
beam-splitter and the polarizing beam-splitter simultaneously in
order to interfere. This implies that the optical paths of the
photons from their point of creation to the beam-splitters must be
accurate to within the coherence length of the photons. Second, if
we take care to use detectors which are not sensitive to the
polarization of the photons, the knowledge that the three
detectors fired at once does not reveal any information regarding
the state of polarization of the photons, thus preserving the
coherence between the two parts of GHZ wave-function. In that
sense, the post-selected state is not a mixed-state, and may be
used in tasks where no further mode-mixing is needed (e.i. GHZ
measurement of violation of local realism). Third, in previous
schemes \cite{zou_02,bouwmeester_99} the input state always
involved some initial entanglement, i.e. entangled pairs of
photons in a bell state $\ket{\psi^-}$. In the new scheme
presented here there is no initial entanglement in the input
state. This fact simplifies considerably the creation of the input
state, and no special setup for creating entangled photons pairs
is needed \cite{kwiat_95,kwiat_99}.

\begin{figure}
    \begin{center}
    \includegraphics[scale=0.27]{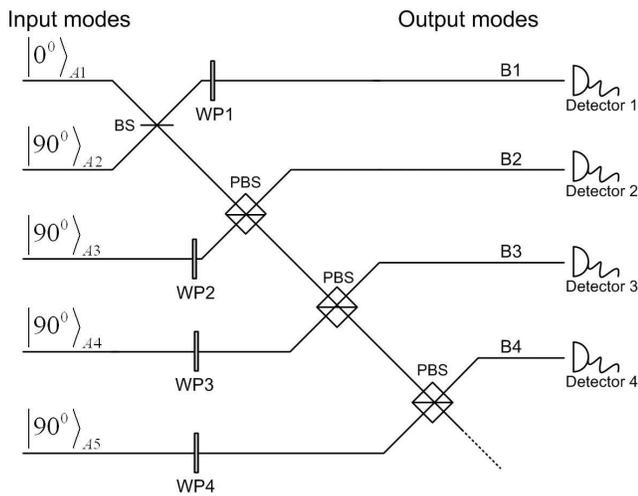}
    \end{center}
    \caption{Generalized n-photons GHZ state creation scheme.}\label{fig3}
\end{figure}

The scheme can be generalized for creating an $n$ photons GHZ
state as defined by
\begin{eqnarray}
\ket{\psi}&=&\frac{1}{\sqrt{2}}\Big{(}\ket{0^o}_{B1}\ket{0^o}_{B2}...\ket{0^o}_{Bn}+\nonumber\\
&&\ket{90^o}_{B1}\ket{90^o}_{B2}...\ket{90^o}_{Bn}\Big{)} \ \ ,
\end{eqnarray}
where $B1..Bn$ are the output modes. The generalized scheme is
depicted in figure \ref{fig3}.
The input state of the generalized scheme consists of $n$-photons
in the non-entangled state
\begin{equation}
\ket{\psi}_{input}=\ket{0^0}_{A1}\ket{90^0}_{A2}\ket{90^0}_{A3}...\ket{90^0}_{An}
\ \ .
\end{equation}
The first two photons are being mixed on a beam-splitter to create
a post-selected $\ket{\psi^-}$ state. Waveplate $WP1$ is used to
flip the polarization of the photon in mode $B1$, just as
explained in the GHZ scheme. $WP2$-$WPN$ are $\lambda/2$
waveplates at $22.5^o$. Thus there is a probabilistic duplication
of the polarization state of one of the photons of the
$\ket{\psi^-}$ state to the photons at modes $B3$ to $BN$ using
the input photons of the modes $A3...An$ and the polarizing
beam-splitters. The probability for the two photons coming from
$A1$ and $A2$ to be in a $\ket{\psi^-}$ state is $1/2$. The
probability of the success of the duplication process is $1/2$ for
each polarizing beam-splitter. Therefore, the success probability
of this scheme is $1/2^{n-1}$.

\begin{figure}
    \begin{center}
    \includegraphics[scale=0.22]{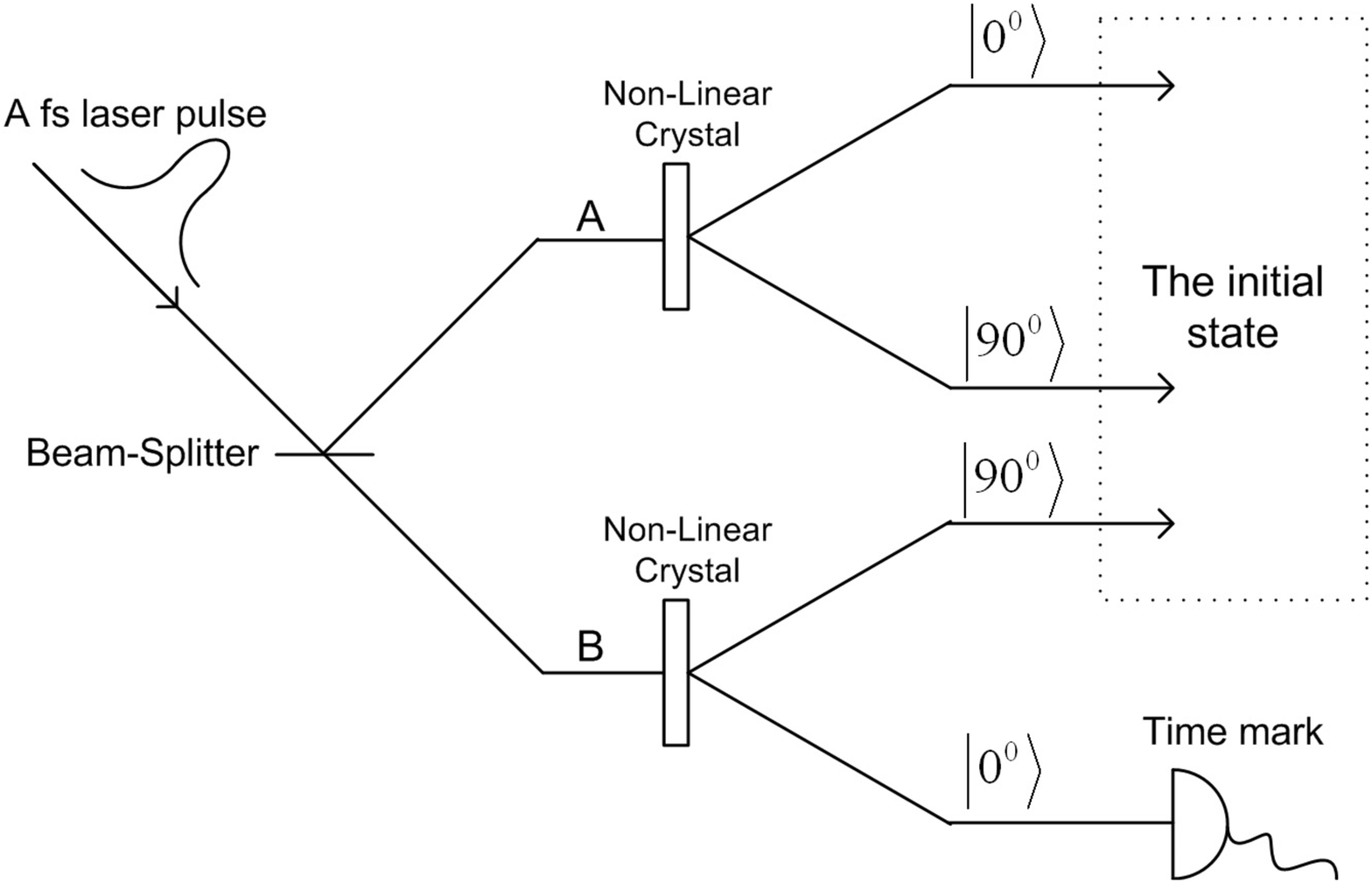}
    \end{center}
    \caption{Input state creation for the $3$-photon GHZ scheme.}\label{fig2}
\end{figure}

The input state can be created using single photon sources.
Nevertheless, the complete technology of these sources is yet to
be established \cite{brunel_99,santori_01}. Naively, one could
suggest the usage of faint continuous lasers as source for the
input state. This cannot be done because of the poissonian
distribution of the laser. To understand why, let us take as an
example two faint sources with poissonian distribution and equal
intensity, entering a beam-splitter from two different directions.
The probability for detecting $n$ photons in a given period of
time $\Delta T$ in each of the modes is given by
\cite{mandel_95_727}
\begin{equation}
p(n)=\frac{W^n e^{-W}}{n!} \ \ ,
\end{equation}
where $W$ is a function of the intensity of the lasers, and the
period $\Delta T$. The event which interest us is when two
distinct photons, each of which is coming from a different source,
interfere at the beam-splitter. The probability for this event to
happen is $p(1)^2$. Nevertheless, there is another undesirable
event in which two photons from one of the sources come at the
same time to the beam-splitter, with probability $2\cdot p(2)$.
This event may cause two detectors placed at the two output modes
of the beam-splitter to record a fire, even though no entanglement
is present in the state. For sources with poissonian distribution
the ratio of the undesired events to the desired ones is $e^{W}$,
and is always greater than $1$ (where $1$ is the limit for sources
with intensity that goes to zero). This means that for poissonian
sources there will be always more ''bad'' events then ''good''
ones, therefore making them not useful to our purpose. Instead, in
figure \ref{fig2} we give a possible way for creating the input
state for the $3$-photons GHZ scheme.
A fs laser pulse with frequency $\omega_p$ undergoes spontaneous
degenerate type II parametric down-conversion (SPDC) in two
non-linear crystals. Thus, two pairs of orthogonally polarized
photons at frequencies $\omega_p/2$ are created. The necessity of
the fs laser pulse is due to the fact that the coherence time of
the photons created in a SPDC process depends on the collection
angle or on the interference filter width, and is of the order of
a few hundred femto-seconds. Therefore, if the creation time
difference between the two photon pairs is not within the
coherence time, they will arrive at the beam-splitters at
different times and will not interfere. A fs laser pulse solves
this problem because it enables us to determine the time of
creation to an accuracy of about $100fs$. The statistical
distribution of pair creation in SPDC is poissonian. This means
that the probability of two pair creation in one of the crystals
is larger than the probability of one pair creation in each of
them. To solve this problem we use the fourth photon as a sign
that a pair was created in the non-linear crystal $B$. Under the
condition that the fourth photon is recorded, the undesired events
are the creation of two pairs in crystal $A$ and of a second pair
in crystal $B$, in addition to a single pair that was created in
crystal $A$ (we neglect undesired events of higher order such as
three pairs creation). The ratio of undesired events to desired
ones equals $W$, where $W$ characterizes the poissonian
distribution of pair creation. This result, in contrast to pure
poissonian sources, shows that by choosing a small enough $W$ we
obtain neglected erroneous counts. Let us give a numerical
example. Choosing $W=10^{-2}$ implies a purity of 99\% in the
state creation. The probability for two pair creation in the two
crystals is $~10^{-4}$, therefore if we use a conventional
Ti:Sapphir laser which operates at a rate of $~80\,MHz$ we obtain
approximately $8000$ double photon pairs per second. Since only
$1/4$ of the times the scheme succeeds, we should expect around
$2000$ GHZ states per second. This rate is five orders of
magnitude higher then the rate obtained in the experiment reported
in Ref. \cite{bouwmeester_99}. Nevertheless, the technique
described here is only useful for small numbers of photons,
because the probability for the simultaneous creation of $N$ pairs
is $(We^{-W})^N$ which is exponentially decreasing with $N$. For
larger $N$ it is necessary to use a true single photon source
which can be triggered synchronically \cite{brunel_99,santori_01}.

In conclusion, in this letter we presented a new scheme for the
probabilistic creation of a GHZ like state of $n$-photons using
only elements of linear optics and single photons detectors. We
also described the input state creation technique which makes use
of non-linear crystals and a fs laser pulse. The new scheme can be
useful in quantum information processing applications where
initial entanglement is needed.

We gratefully acknowledge helpful discussions with Moshe Shuker,
Amit Ben-Kish, Ofer Firstenberg, Amnon Fisher and Amiram Ron.


\end{document}